\documentclass[preprint,showpacs,showkeys]{elsarticle}
\usepackage[T1]{fontenc}
\usepackage{bm}
\usepackage{graphicx}
\usepackage{amssymb}
\usepackage{amsfonts}
\usepackage{amsmath}
\usepackage{multirow}
\usepackage{pstricks}
\usepackage{graphicx}

\begin{document}

\title{MBE growth and characterization of a II-VI distributed Bragg reflector and microcavity lattice-matched to MgTe}

\author{J.-G. Rousset}

\author{J. Kobak}

\author{T. S{\l}upinski}

\author{T. Jakubczyk}

\author{P. Stawicki}

\author{E. Janik}

\author{M. Tokarczyk}

\author{G. Kowalski}

\author{M. Nawrocki}

\author{W. Pacuski}

\address{Institute of Experimental Physics, Faculty of Physics, University of Warsaw, ul. Ho\.za 69, PL-00-681 Warszawa, Poland}



\begin{abstract}

We present the realization and characterization of a 20 fold, fully lattice-matched epitaxial distributed Bragg reflector based on (Cd,Zn)Te and (Cd,Zn,Mg)Te layers. We also present a microcavity based on (Cd,Zn,Mg)Te containing a (Cd,Zn)Te quantum well. Reflectivity spectra, photoluminescence in real space and in far field are presented.

\end{abstract}

\begin{keyword}
Molecular beam epitaxy \sep Cadmium compounds \sep Zinc compounds \sep Tellurides \sep Semiconducting II-VI materials \sep Magneto-optic materials
\PACS 78.20.Ci \sep 78.67.Pt \sep 68.37.Hk \sep 78.55.Et

\end{keyword}




\maketitle


\section{Introduction}

In III-V semiconductor compounds, the lattice-matching between GaAs and AlAs allows for the growth of monolithic structures and to realize devices for optoelectronics  \cite{Gayral_1998,Dousse_2010}. For II-VI compounds, the challenge in designing and realizing epitaxial distributed Bragg reflectors (DBRs) and microcavities relies on the compromise between opposite constrains: the lattice-matching and a high refractive index contrast. Lattice-matched DBRs based on II-VI compounds have been reported so far for four substrates (or buffers):  GaAs  \cite{Kruse_2004}, InP  \cite{Guo_2000}, ZnTe  \cite{Pacuski_2009}, and (Cd,Zn)Te \cite{LeSiDang_1998,Kasprzak_2006}. For this last one, the substrate was Cd$_{0.88}$Zn$_{0.12}$Te and the DBR was based on (Cd,Mg)Te and (Cd,Mn)Te layers grown on the (Cd,Zn)Te buffer. In such a structure, the presence of Mn, which has strong magnetic properties, in the DBRs layers can be a disturbance if one wants to explore magnetooptical effects related to quantum wells  \cite{Ulmer_1996,Brunetti_2005,Brunetti_2006}. We present here the realization of unstrained DBR lattice-matched to MgTe, based on the stack of (Cd,Zn,Mg)Te layers with various Mg contents. Enhancements in the design of the DBR led us to the realization and characterization of a monolithic, optically active microcavity containing a lattice-matched Cd$_{0.86}$Zn$_{0.14}$Te  quantum well (QW) surrounded by good quality DBRs.

\begin{figure}[!h]
  \includegraphics[width=0.7\linewidth]{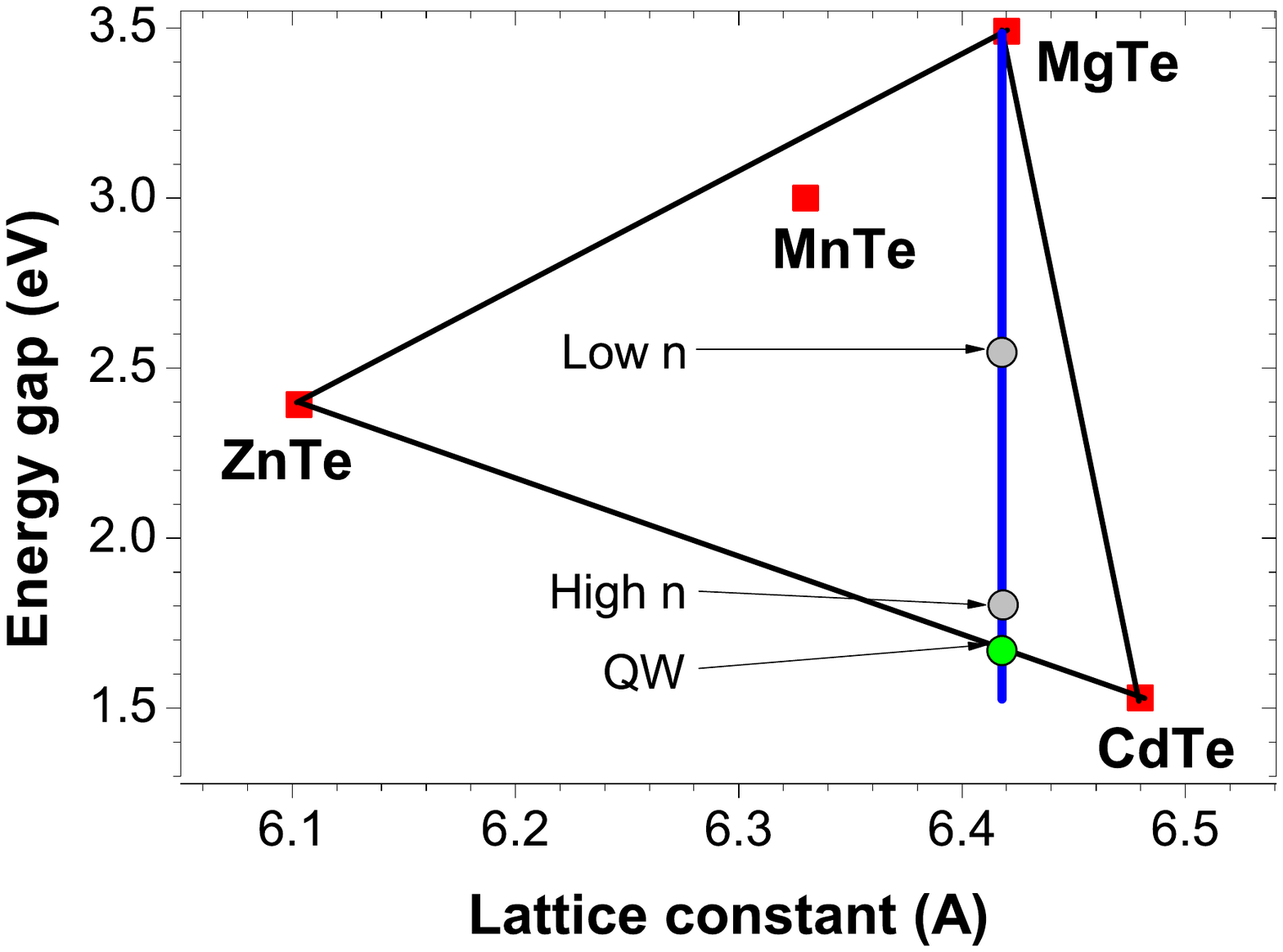}\\
\centering  \caption{The choice of having a structure lattice-matched to MgTe allows to tune the refractive index contrast through Mg content of the layers without changing the lattice constant. Line: MgTe lattice constant - dots: high and low refractive index layers, QW.}
  \label{latmatch}
\end{figure}

As shown in Fig. \ref{latmatch}, our choice of a structure lattice-matched to MgTe is justified by the possibility of tuning the refractive index and the band gap of the layers independently on the lattice parameter, depending only on the Mg content. With this concept, the whole structure including DBRs and QW can be lattice-matched.

The DBR, microcavity and QW were grown by molecular beam epitaxy (MBE) technique. The MBE machine provided by SVT Associates consists of two growth chambers (for III-V and II-VI semiconductor compounds), each equipped with a reflectivity setup for in situ measurements. Monitoring the reflectivity spectra during the growth allows us to verify the growth rate and thickness of the layers, which is a crucial parameter for the growth of microcavities or DBRs \cite{Kruse_2002, Shenk_2002,Mizutani_2006,Biermann_2011}.

The structures are grown on a GaAs:Si substrate (100) at the temperature of $346^{\circ}$. A Cd$_{0.86}$Zn$_{0.14}$Te buffer about 800~nm thick is grown to relax the strain due to the lattice mismatch with GaAs. The lattice parameter and the composition of the buffer was estimated in-situ from streaks spacing on the RHEED (reflection high energy electron diffraction) pattern.

\section{Distributed Bragg Reflector}

\begin{figure}
\centering  \includegraphics[width=0.7\linewidth]{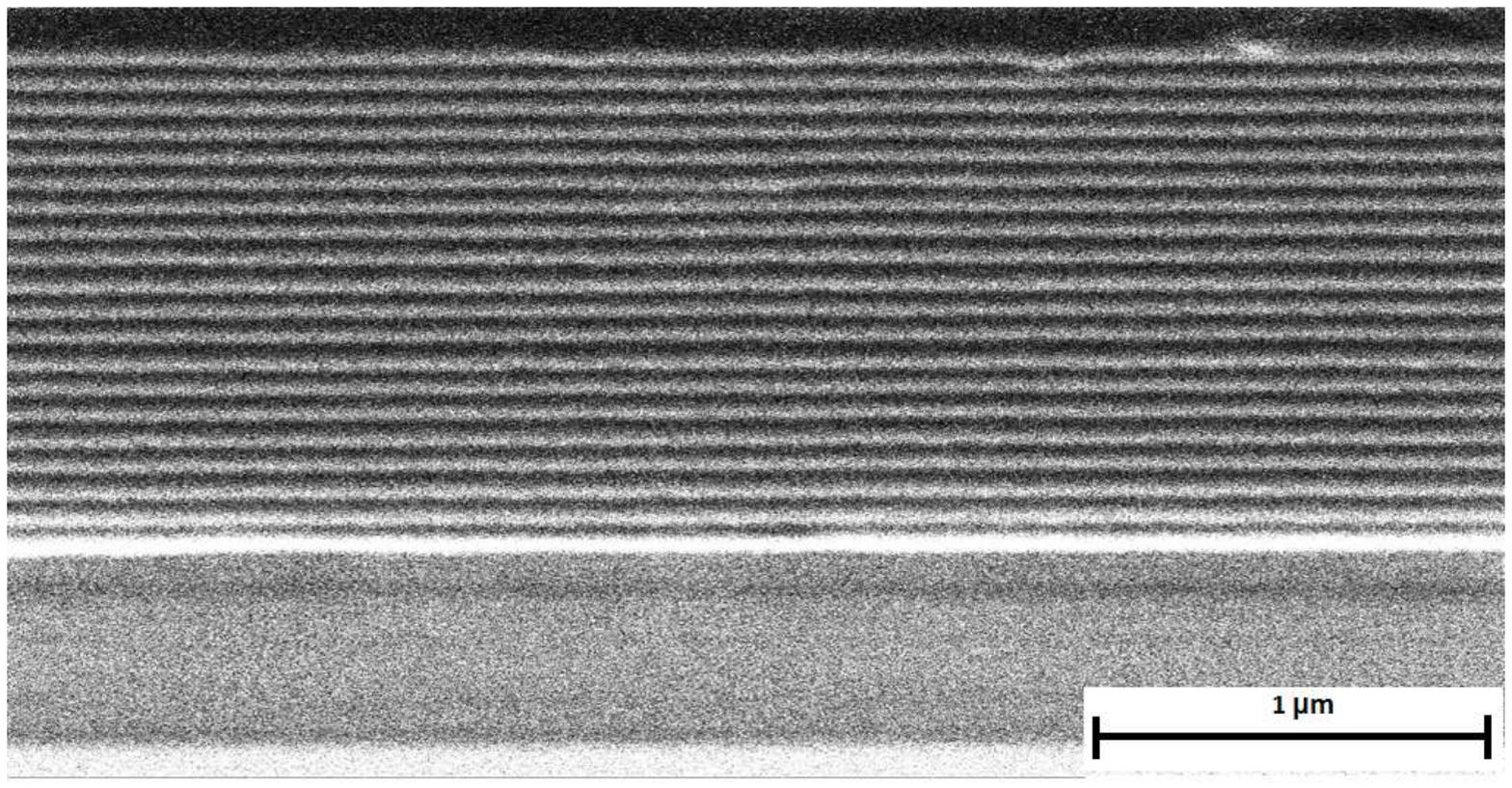}\\
  \caption{SEM imaging of the DBR.}
  \label{imDBR}
\end{figure}

The DBR containing 20 pairs of low and high refractive index layers was designed so that the center of the stopband is at a wavelength of $\lambda _0\approx 900$~nm. The layers of the DBR are made of $\lambda/4$n layers of Cd$_{0.86}$Zn$_{0.14}$Te for the high refractive index and a 20 fold Cd$_{0.86}$Zn$_{0.14}$Te/MgTe superlattice for the low refractive index. The growth times of the superlattice layers are set to have an effective concentration of 50\% Mg in the resulting digital alloy. The structure grown was observed by scanning electron microscope (SEM) imaging. As presented in Fig. \ref{imDBR} the layers thicknesses are regular and the interfaces are smooth.

\begin{figure}[!h]
  \includegraphics[width=0.9\linewidth]{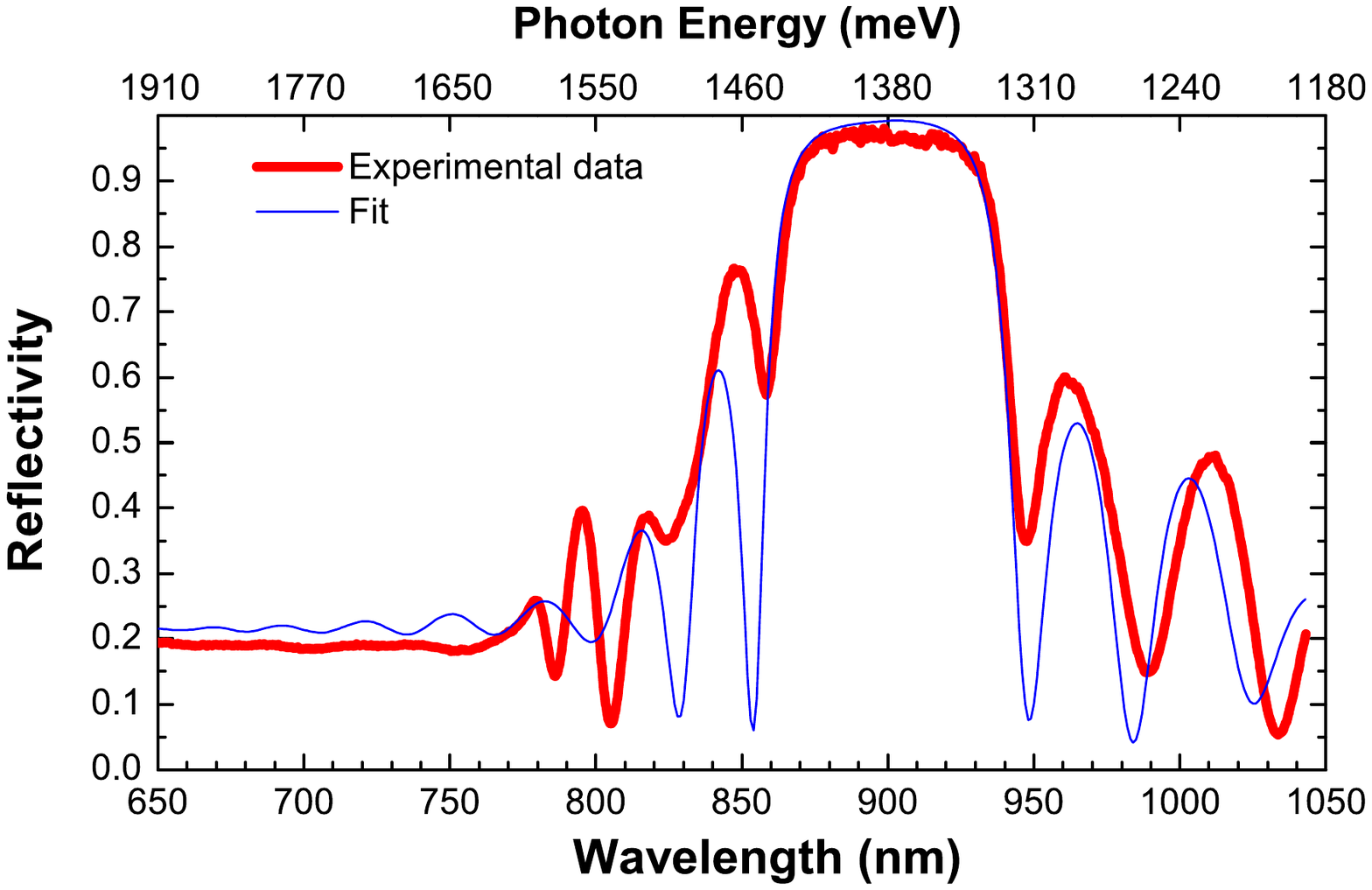}\\
\centering  \caption{Reflectivity spectra: measurement and simulation at room temperature. The maximum reflectivity is above 95\%, the stopband is $\approx 70$~nm wide giving a refractive index contrast $\Delta n \approx 0.34$.}
  \label{reffitDBR}
\end{figure}

The results from reflectivity measurements compared to the simulation by the transfer matrix method  \cite{Yeh_1988} are presented in Fig. \ref{reffitDBR}. The DBR exhibit a maximum reflectivity above 95\%. The stopband width measured is $\approx 70$~nm which allows us to evaluate the refractive index contrast using the formula \cite{Yeh_1988} given in eq. \ref{width}, where $\Delta \lambda$ sets for the stopband width, $\lambda_0$ the center of the stopband,  $\tilde n$ the average refractive index.

\begin{equation}\label{width}
    \frac{\Delta \lambda}{\lambda_0}=\frac{4}{\pi} sin^{-1} (\frac{|n_2-n_1|}{n_1+n_2}) \approx \frac{2}{\pi} \frac{\Delta n}{\tilde n}
\end{equation}

Considering that the average refractive index \cite{Marple_1964} is $\tilde n \approx 2.8$, the refractive index contrast is $\Delta n \approx 0.34$.

\begin{figure}[!h]
  \includegraphics[width=0.8\linewidth]{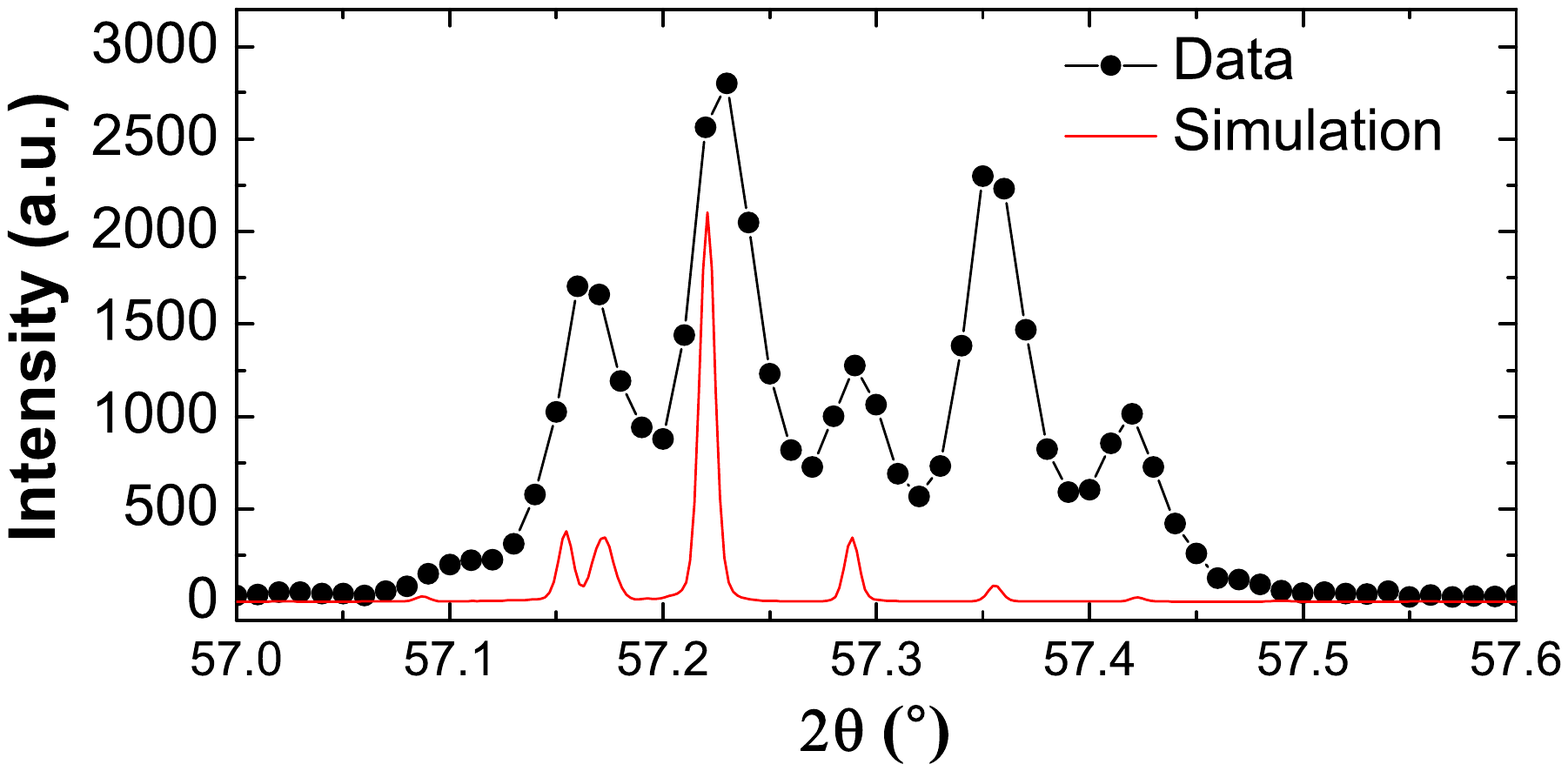}\\
\centering    \caption{X-ray diffraction measurement (dots) and simulation (solid line) assuming that the structure is relaxed and lattice-matched to MgTe.}
  \label{Xray}
\end{figure}

X-ray diffraction measurements and simulation show that the whole structure (except the substrate) is matched to the lattice constant of MgTe \cite{Dynowska_1999} (lattice mismatch $\Delta a/a <0.3\%$) (see Fig. \ref{Xray}).

\begin{figure}[!h]
  \includegraphics[width=0.8\linewidth]{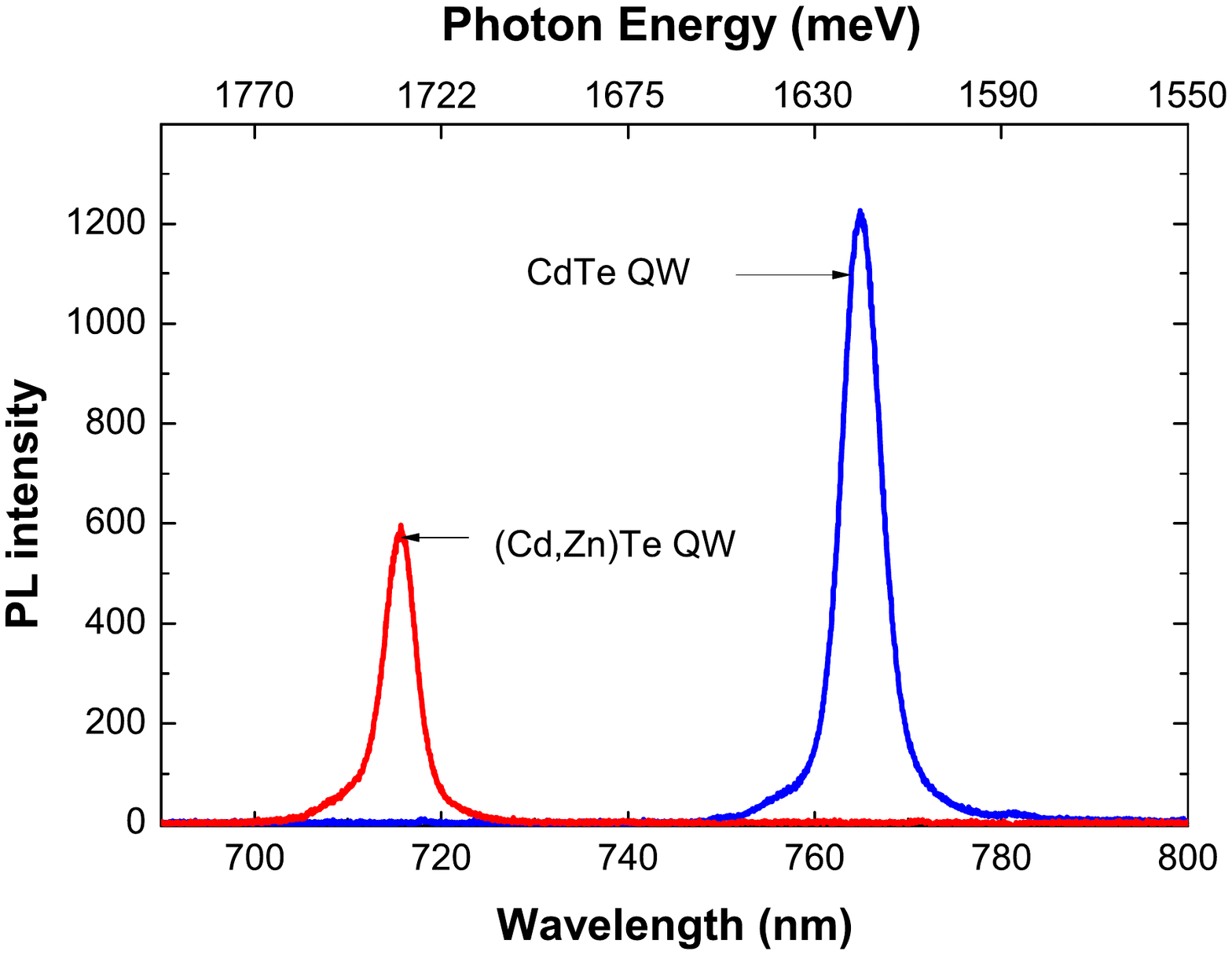}\\
\centering    \caption{Photoluminescence of CdTe QW (strained) and unstrained Cd$_{0.86}$Zn$_{0.14}$Te QW. In both cases the QW is $10$~nm thick, the barrier is Cd$_{0.77}$Zn$_{0.13}$Mg$_{0.1}$Te and the width  (FWHM) of PL peak is about $10$~meV ($4$~nm).}
  \label{QWs}
\end{figure}

\section{Quantum wells}

In order to obtain a fully lattice-matched microcavity, we designed a Cd$_{0.86}$Zn$_{0.14}$Te QW with Cd$_{0.77}$Zn$_{0.13}$Mg$_{0.1}$Te barriers.  Fig. \ref{QWs} shows photoluminescence spectrum of this ternary QW measured at liquid helium temperature and compared to the photoluminescence spectrum of a CdTe QW.

The PL amplitude of the ternary Cd$_{0.86}$Zn$_{0.14}$Te QW is twice smaller than for the strained CdTe QW, however the half peak width is about $10$~meV  ($4$~nm)for both QWs.

\section{Microcavity with quantum well}

\begin{figure}[!h]
  \includegraphics[width=1\linewidth]{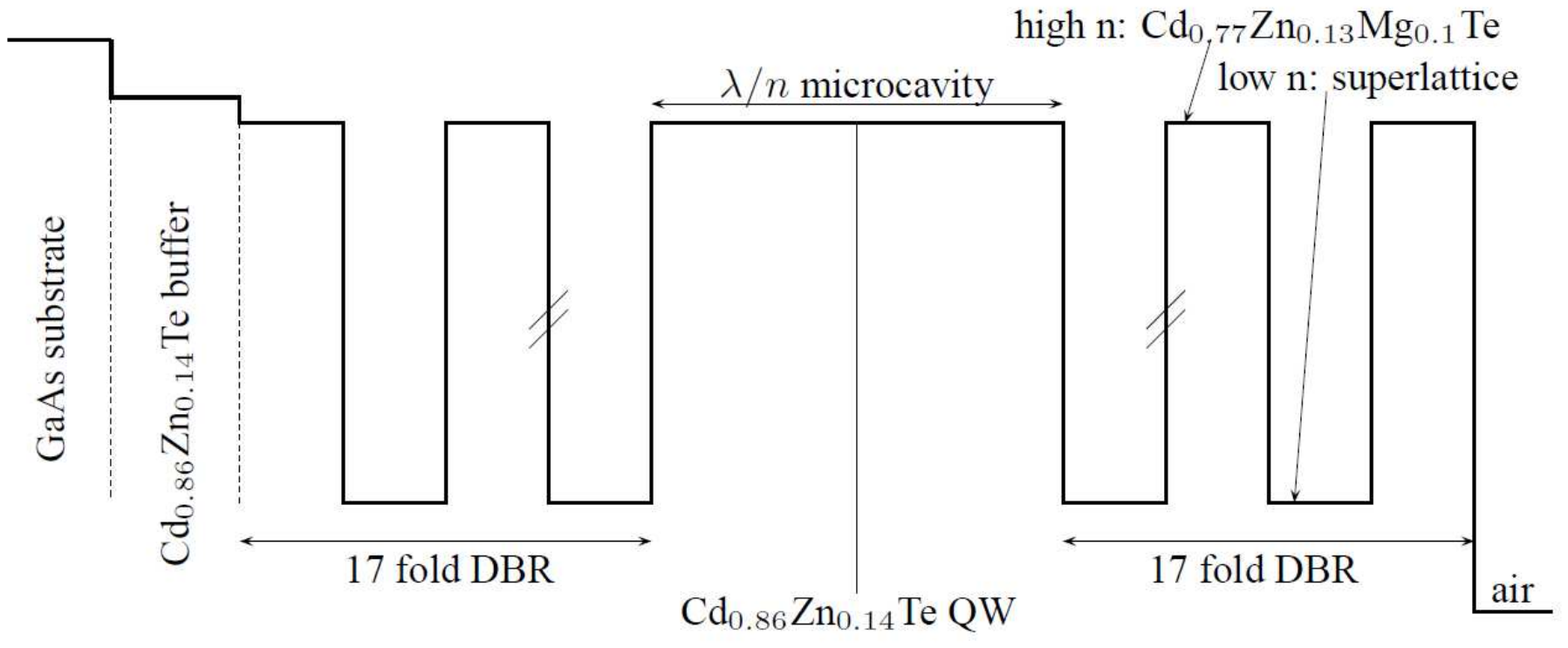}\\
\centering    \caption{Structure of the microcavity. Vertical axis: refractive index, horizontal axis: depth of the sample.}
\label{microcav}
\end{figure}

In order to obtain a fully lattice-matched microcavity embedding an optically active QW, the design of the DBR previously presented was slightly modified. The thicknesses of the layers are designed so that the center of the stopband is at $\lambda _0 = 720$~nm corresponding to the emission of the Cd$_{0.86}$Zn$_{0.14}$Te QW. In addition, the insertion of the optically active Cd$_{0.86}$Zn$_{0.14}$Te QW requires increasing the energy gap of the high refractive index layer so that it also plays the role of the barrier for the QW and does not absorb the QWs emission. This is achieved by adding a small amount of Mg in the high refractive index layer: Cd$_{0.77}$Zn$_{0.13}$Mg$_{0.1}$Te. In order to keep a relatively high refractive index contrast, the effective Mg content in the low refractive index layer (superlattice) has also been increased close to $60\%$. As presented in Fig. \ref{microcav}, the whole structure consists of a $2$0~nm thick QW at the center of a $\lambda /n$ microcavity embedded between two 17 fold DBRs.

\begin{figure}[h]
  \includegraphics[width=1\linewidth]{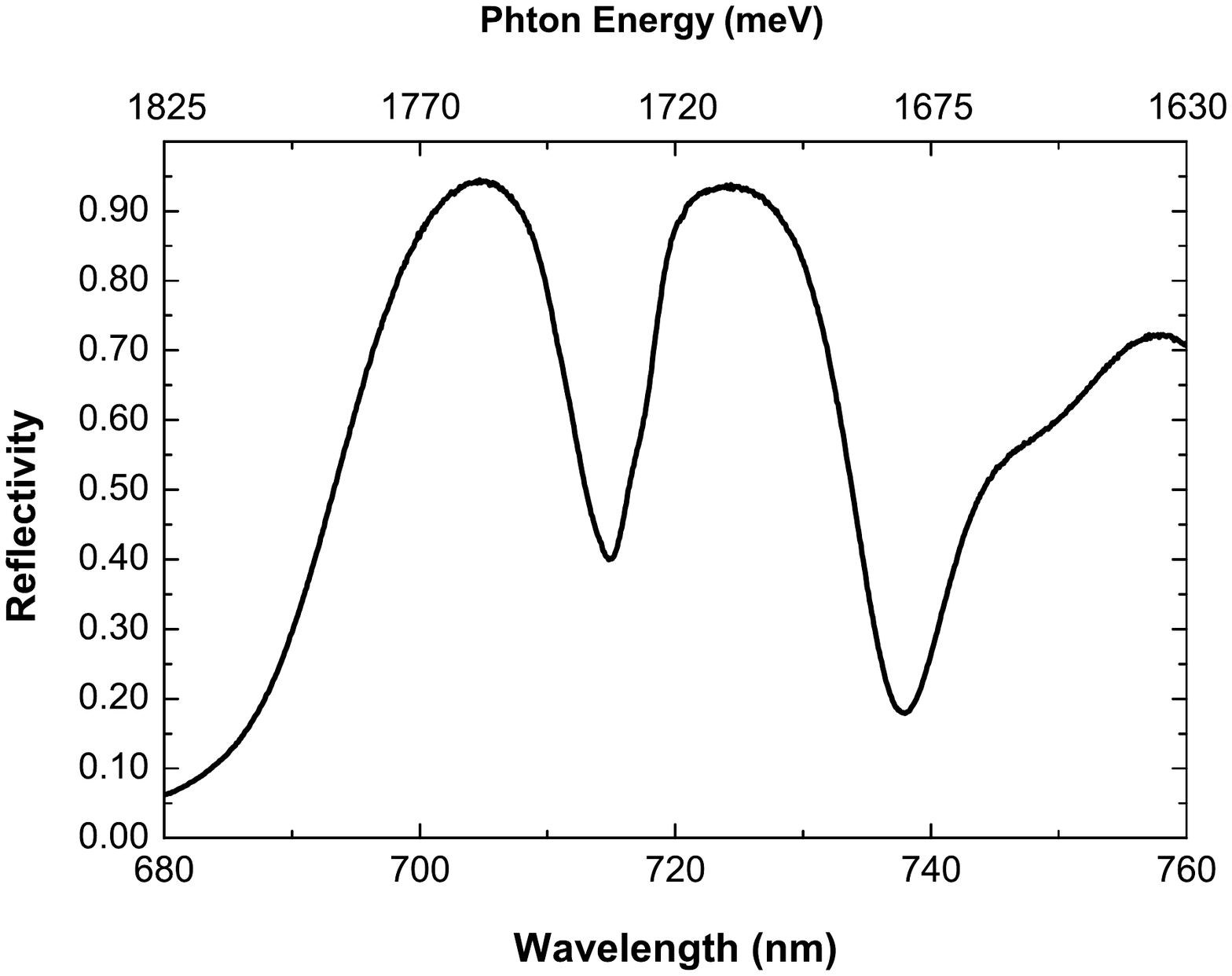}\\
\centering  \caption{Reflectivity spectrum of the microcavity measured at $T=8~K$. The quality factor of the cavity is $Q\approx 150$}
  \label{refmicrocav}
\end{figure}

Reflectivity and photoluminescence spectra have been measured at $T=8$~K. As shown in Fig. \ref{refmicrocav}, the microcavity exhibit a maximum reflectivity of about $90\%$. The quality factor is  $ Q=\frac{\lambda}{\Delta \lambda}\approx 150$.


\begin{figure}[!h]
  \includegraphics[width=1\linewidth]{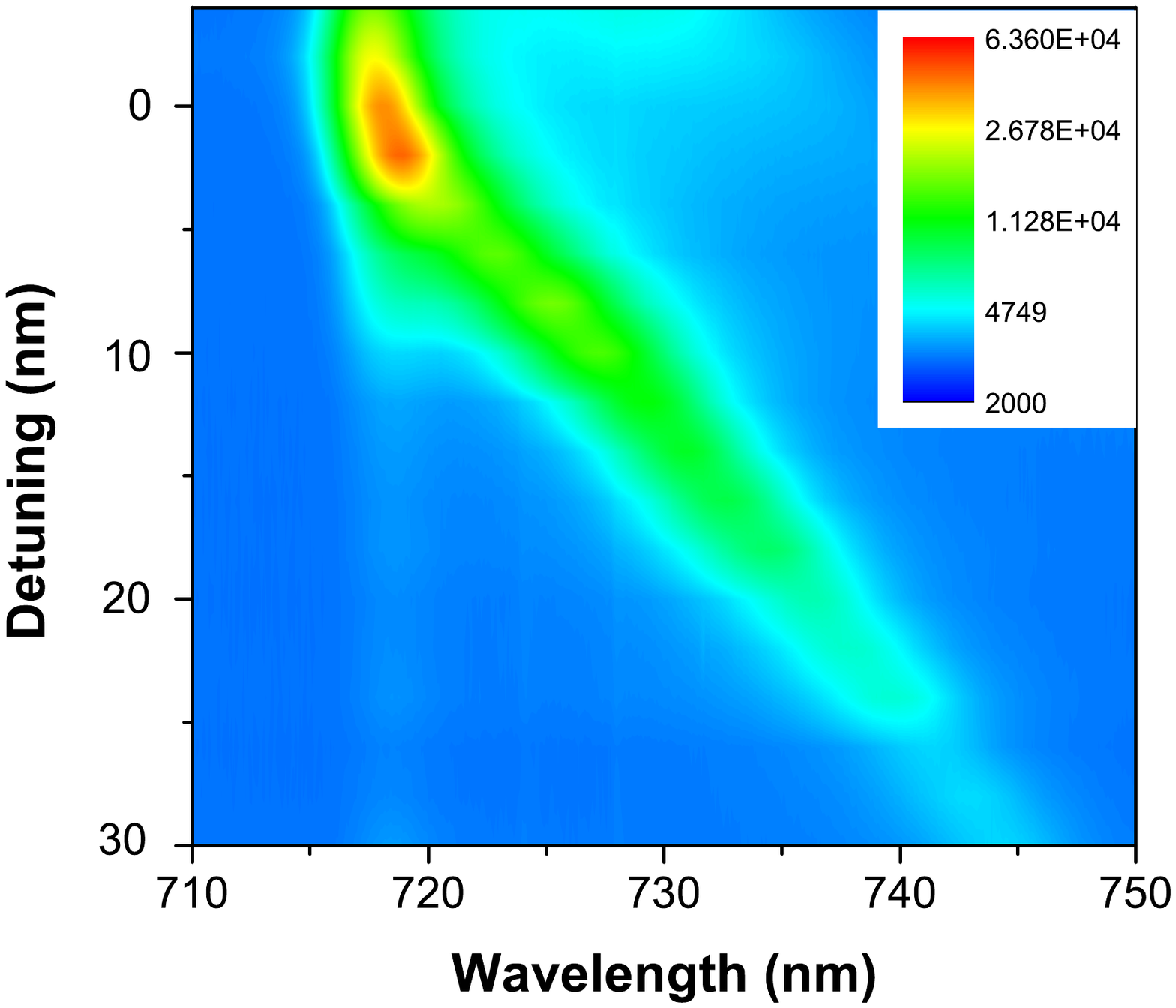}\\
\centering    \caption{Map showing the photoluminescence as a function of the wavelength for various values of the detuning.}
  \label{mapPL}
\end{figure}

The photoluminescence measurements presented in Fig. \ref{mapPL} show the tuning of the cavity mode to the emission of the QW resulting in the enhancement of the emission intensity at $\lambda_0 = 716$~nm, in agreement with our expectation. Different values of the detuning are obtained for different places on the sample. Indeed, the sample was not rotated during growth which results in a gradient of the layers' thicknesses.

\begin{figure}[h]
  \includegraphics[width=1\linewidth]{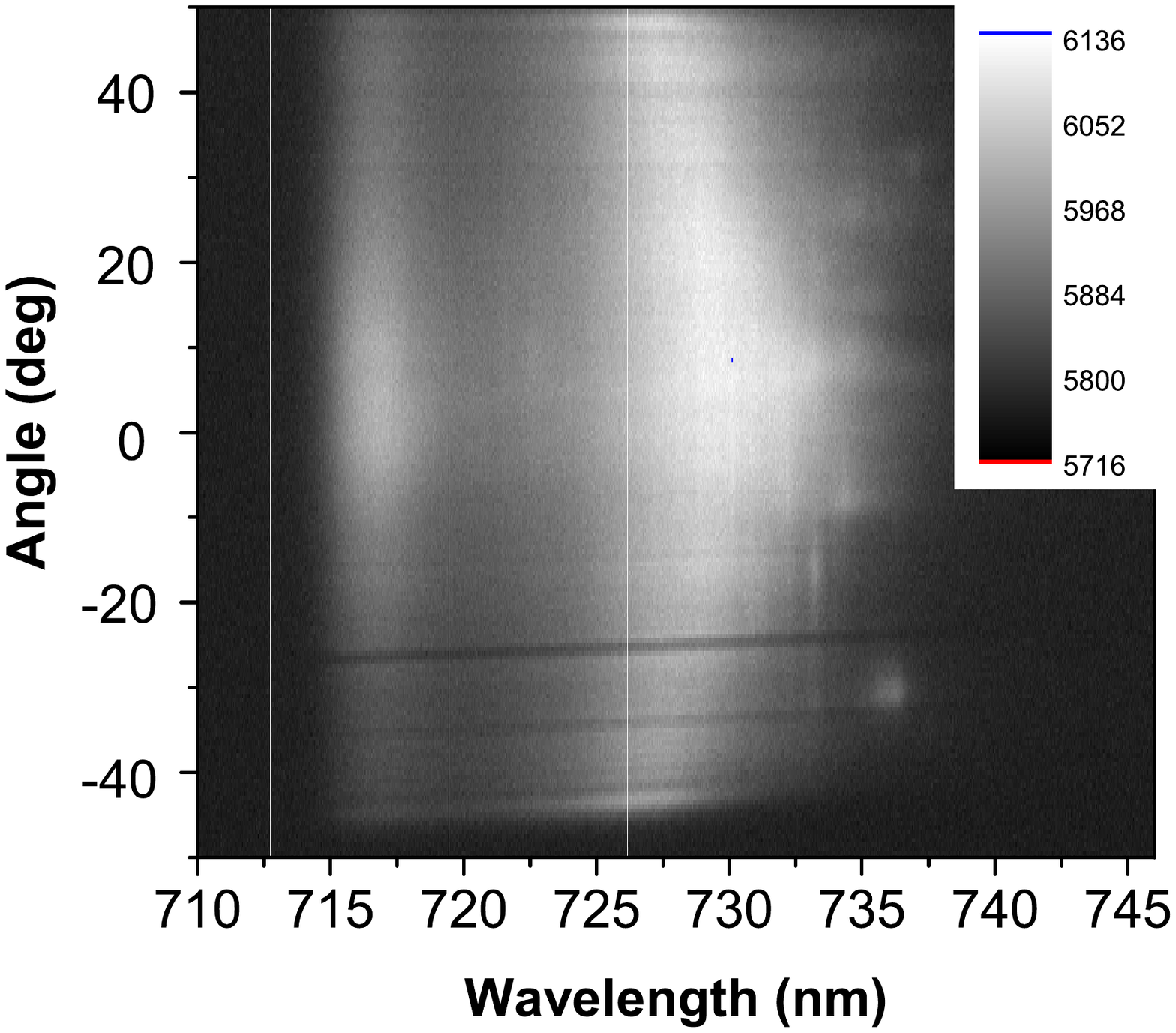}\\
\centering    \caption{Far field emission. The wavelength of the cavity mode depends on the propagation direction of the light. The emission from the QW remains at the same wavelength.}
  \label{farfield}
\end{figure}

Far field PL measurements were also conducted (Fig. \ref{farfield}) and show the dependence on the angle of the emission from the cavity whereas the emission from the QW remains unchanged.  It shows that in the a microcavity, exciton dispersion is much weaker than photon dispersion.  This measurement was realized using the Fourier plane imaging method  \cite{Jakubczyk_CEJP2011, Kasprzak_2006}. By using this technique, adding more quantum wells in the cavity, having QW lines more sharp,  increasing quality factor of the microcavity, and using high excitation power we expect to observe Bose-Einstein condensation of exciton polaritons \cite{Kasprzak_2006}. In present study we are very far from this conditions.

\section{Conclusion}

The growth of DBR, QW and microcavity lattice-matched to MgTe based on a Cd$_{0.86}$Zn$_{0.14}$Te buffer was presented. This allowed the tuning of the refractive index and energy gap through the Mg content of the layers. The 20 fold DBR exhibits a reflectivity above $95\%$. X-ray diffraction measurements show that the whole structure is lattice-matched to MgTe. The tuning of the emission from the  Cd$_{0.86}$Zn$_{0.14}$Te QW to the cavity mode results in an enhancement of the photoluminescence at $\lambda_0 = 716$~nm. The far field emission shows the dependence of the emission from the cavity on the angle of the propagating light. The realization of such a microcavity is a first step toward the realization of a vertical cavity surface emitting laser (VCSEL).

\section*{Acknowledgement}

This work was supported by Polish public funds in years 2011 - 2014 (NCBiR project LIDER). Research was carried out with the use of CePT, CeZaMat and NLTK infrastructures financed by the European Union - the European Regional Development Fund within the Operational Programme "Innovative economy" for 2007-2013.

\end{document}